\documentclass[preprintnumbers,superscriptaddress,showpacs,pra]{revtex4}%
\usepackage[colorlinks=true,linkcolor=blue]{hyperref}
\usepackage{amssymb}
\usepackage{amsfonts}
\usepackage{amsmath}
\usepackage{graphicx}%
\providecommand{\U}[1]{\protect\rule{.1in}{.1in}}

\let\stdsection\section
\renewcommand\section{\nopagebreak\stdsection}
\begin{document}
\title{Geometric momentum for a particle constrained on a curved hypersurface}
\author{Q. H. Liu}
\affiliation{School for Theoretical Physics, and Department of Applied Physics, Hunan
University, Changsha, 410082, China. Tel/fax 86-731-88820378/86-731-88822332,
Email: quanhuiliu@gmail.com}

\begin{abstract}
A strengthened canonical quantization scheme for the constrained motion on
curved surface is proposed with introduction of the second category of
fundamental commutation relations between Hamiltonian and positions/momenta,
whereas those between positions and moments are categorized into the first. As
an $N-1$ ($N\geq2$) dimensional surface is embedded in an $N$ dimensional
Euclidean space, we obtain the geometric momentum $\mathbf{p}=-i\hbar
(\mathbf{\nabla}_{S}+M\mathbf{n}/2)$ where $\mathbf{\nabla}_{S}$ denotes the
gradient operator on the surface and the $M\mathbf{n}$ is the mean curvature
vector. For the surface is the spherical one of radius $r$, we resolve in a
lucid and unambiguous manner a long-standing problem of the geometric
potential that proves to be $V_{g}=(N-1)(N-3)\hbar^{2}/(8mr^{2})$.

\end{abstract}
\date{\today}

\maketitle
\preprint{REV\TeX4-1 }

\section{Introduction}

The quantum mechanics for a non-relativistic particle that is constrained to
remain on a curved hypersurface attracts much attention
\cite{weinberg,jk,dacosta,jpn1,jpn2,jpn3,liu07,EM,JD,OB,SOintl,liu11,exp1,exp2}%
. As well-known, Dirac's quantum theory for a constrained motion
\cite{dirac1,dirac2} does not always produce physically significant results
and usually exhibits certain difficulties in application \cite{Klauder}. For
instance, we do not have a well-defined form for either momentum or
Hamiltonian after quantization \cite{jpn1,jpn2,liu11}. In this paper, we
propose a strengthened canonical quantization scheme (SCQS) for the
constrained motion on the surface. Explicitly, we deal with an $N-1$ ($N\geq
2$) dimensional hypersurface $S^{N-1}$ and its equation is either
$f(\mathbf{x})=0$ or parametric form $\mathbf{x}=\{x_{i}(u)\}$ in the $N$
dimensional flat Euclidean space $R^{N}$, where $x_{i}$\text{ (}%
$i,j,k,\ell=1,2,3,\ldots,N$) stand for the Cartesian coordinates and $u^{%
\mu
}$ ($%
\mu
,\nu=1,2,3,\ldots,N-1$) symbolize the local coordinates on the surface.
According to Dirac, such an constraint belongs to the second kind
\cite{dirac2}. In our approach, we do not quantize the local coordinates $u^{%
\mu
}$ and corresponding momenta $p_{\mu}$, whereas we treat $u^{%
\mu
}$ as parameters and quantize the Cartesian coordinates $x_{i}$ and its
corresponding momentum $p_{i}$.

At first, let us briefly review and comment on the so-called confining
potential technique that leads to the well-defined \textit{geometric potential
\ }\cite{note}. For a particle constrained on the surface, we can establish an
effective theory in the following way. First, to formulate the Schr\"{o}dinger
equation in $R^{N}$, explicitly in a curved shell of an equal and finite
thickness $\delta$ along normal direction $\mathbf{n}$, and let\ the
intermediate surface of the shell coincide with the prescribed one $S^{N-1}$.
And the particle moves within the range of the same width $\delta$ due to a
confining potential across the surface along the normal direction $\mathbf{n}%
$, such as one-dimensional parabolic one or simply the square potential well.
Second, to take the limit $\delta\rightarrow0$, we have an effective kinetic
energy operator different from the well-known one $-\hbar^{2}/(2m)\nabla
_{LB}^{2}$ as \cite{jpn2},
\begin{equation}
-\frac{\hbar^{2}}{2m}\nabla_{LB}^{2}\rightarrow-\frac{\hbar^{2}}{2m}\left(
\nabla_{LB}^{2}+v_{g}\right)  ,\text{ }v_{g}=\frac{1}{4}\left(  2Tr(\mathbf{k}%
)^{2}-(Tr\mathbf{k})^{2}\right)  , \label{GP}%
\end{equation}
where $V_{g}\equiv$ $-\hbar^{2}/(2m)v_{g}$ is the curvature-induced\textit{
}potential that is usually called as the geometric potential,\ and $v_{g}$ is
purely determined by the principal curvatures $\mathbf{k}$ \cite{jpn2}. This
approach seems to suffer from a theoretical shortcoming: we do not know why
such an establishment of the quantum theory can not be directly on the
surface. If so, it predicts a vanishing geometric potential that would
contradict the recent experiments \cite{exp1,exp2}. In fact, Dirac's canonical
quantization procedure simply excludes such an attempt, as we see shortly.

Next, let us recall of two differential geometric facts for hypersurfaces: 1,
The definition of the mean curvature $M=Tr\mathbf{k=-\partial}n_{j}/\partial
x_{j}$ (rather than a true average $Tr\mathbf{k/(}N-1\mathbf{)}$ which is also
widely used) \cite{NYgeometry,ESgeometry,JNgeometry,geometry}, and the mean
curvature vector $M\mathbf{n}$ which satisfies ${\nabla_{S}\cdot}%
\mathbf{n=-}M$ \cite{NYgeometry,ESgeometry,JNgeometry,geometry}, where the
surface gradient ${\nabla_{S}}$ is defined by the difference of the usual
gradient $\nabla_{N}$ in $R^{N}$ and its component along the normal direction
$\mathbf{n}\partial_{n}$: ${\nabla_{S}\equiv}\left(  {\partial}%
\mathbf{x/\partial}u^{%
\mu
}\right)  \partial^{\mu}\equiv\mathbf{e}_{i}(\delta_{ij}-n_{i}n_{j}%
)\partial_{j}$ $=\nabla_{N}-\mathbf{n}\partial_{n}$ \cite{JNgeometry} with
$\mathbf{e}_{i}$ being the unit vector of the $i$th Cartesian coordinate and
$\delta_{ij}-n_{i}n_{j}$ being the orthogonal projection from $R^{N}$ to the
plane tangential to the $S^{N-1}$. 2, The Laplace-Beltrami operator
$\nabla_{LB}^{2}$ is given by $\nabla_{LB}^{2}={\nabla_{S}\cdot\nabla_{S}}$
$=\partial_{i}(\delta_{ij}-n_{i}n_{j})\partial_{j}=\Delta_{N}+M\partial
_{n}-\partial_{n}^{2}$ with $\Delta_{N}\equiv\partial_{i}\partial_{i}$ the
usual Laplacian operator \cite{NYgeometry,ESgeometry,JNgeometry,geometry}, and
$\nabla_{LB}^{2}\mathbf{x=}M\mathbf{n}$
\cite{NYgeometry,ESgeometry,JNgeometry,geometry}. It is evident that
$\nabla_{LB}^{2}$, ${\nabla_{S}}$, $\mathbf{n,}$and $M\mathbf{n}$ are all
geometric invariants. The Einstein summation convention for repeated indices
is used throughout the paper.

The organization of the paper is as what follows. In section II, the SCQS is
proposed, and as a consequence the geometric momentum is given In section III.
In section IV, by use of the scheme we deal with the geometric potential for a
particle on an $N-1$ dimensional spherical surface and thus resolve a
long-standing and highly controversial problem on the form of the geometric
potential. Also in section IV, we show that the motion on the surface
possesses a dynamical $SO(N,1)$ group symmetry. In section V, we finally
conclude and remark the present approach.

\section{Second category of fundamental commutation relations}

In our SCQS, there are two categories of fundamental commutation relations
(FCR). The existent one is classified into the first one, which is for
positions and momenta ($x_{i}$, $p_{i}$): $[x_{i},x_{j}]$, $[x_{i},p_{j}]$,
and $[p_{i},p_{j}]$. When constraints are released, these FCR reduce to
$[x_{i},p_{j}]=i\hbar\delta_{ij}$ and all other commutators vanishing. This is
the fundamental postulate for positions and momenta in quantum mechanics,
proposed by Dirac \cite{dirac1,dirac2}, without requiring them to be
Cartesian. It holds true universally if applicable, but is not so practical
for, e.g., a system that does have a classical analogue. For sake of the
practicality, Dirac immediately developed his FCR with an additional
hypothesis that \textit{the Hamiltonian is the same function of the canonical
coordinates and momenta in the quantum theory as in the classical theory,
provided that the Cartesian coordinates must be used }\cite{dirac1}\textit{.}
As a consequence, the equation of motion $dO/dt=[O,H]_{D}$ for an observable
$O$ remains the same form in quantum theory $dO/dt=(i/\hbar)[H,O]$
\cite{1925}, where $[O_{1},O_{2}]_{D}$ denotes the Dirac bracket in general
which includes the Poisson one as the special case. Since then, some,
including us, takes for granted that the Cartesian coordinates in the
underlying flat Euclidean space must be used in performing the canonical
quantization, and the existence of the space is considered a fundamental
postulate in non-relativistic quantum mechanics
\cite{weinberg,dirac1,Klauder,Schiff,Greiner,essen}.

For a system with constraints of the second kind, Dirac's canonical
quantization procedure needs to be further strengthening for sake of the
practicality. From this category of the FCR, there are many forms of the
momentum $\mathbf{p}$ and if substituting these forms of $\mathbf{p}$ into
Hamiltonian such as $H=p^{2}/2m+V$, there are various forms of Hamiltonian in
quantum mechanics. It is then possibly to hypothesize that the forms of the
momentum and the Hamiltonian are determined by imposing the algebraic
structure between observables $\mathit{[O}_{1}\mathit{,O}_{2}\mathit{]}_{D}$
that preserves in quantum mechanics to the extent possible, i.e., in quantum
mechanics $\mathit{[O}_{1}\mathit{,O}_{2}\mathit{]}\equiv i\hbar
\mathit{[O}_{1}\mathit{,O}_{2}\mathit{]}_{D}$. Fundamentally, we introduce the
secondary category of FCR as $[\mathbf{x,}H]\equiv i\hbar\lbrack
\mathbf{x},H]_{D}$ and $[\mathbf{p,}H]\equiv i\hbar\lbrack\mathbf{p},H]_{D}$.

Two observations follows. 1, For the systems without any constraints, the
secondary category of FCR is automatically satisfied. 2, For an intrinsic
description of a hypersurface, the global Cartesian coordinate system does not
exist and even the local one $\left\{  u^{%
\mu
}\right\}  $ can be used approximately. Because the local coordinates
$\left\{  u^{%
\mu
}\right\}  $ must never be Cartesian, the secondary category of FCR $[u^{%
\mu
}\mathbf{,}H]\equiv(i\hbar)[u^{%
\mu
},H]_{D}$ and $[p_{\mu}\mathbf{,}H]\equiv(i\hbar)[p_{\mu},H]_{D}$ would hardly
be all satisfied, as illustrated by \cite{liu11,arXiv1,arXiv2}. So, the
fundamental importance of the Cartesian coordinates within the canonical
quantization procedure excludes the possibility of an attempt to get the
proper form of the quantum Hamiltonian by means of a direct quantization of
the local coordinates $\left\{  u^{%
\mu
}\right\}  $ and their generalized momentum $\left\{  p_{\mu}\right\}  $
\cite{dirac1,Schiff,Greiner,essen}. As a consequence of this observation, we
resort to $R^{N}$ to deal with $S^{N-1}$ that is embedded in $R^{N}$ to
perform the canonical quantization.

For the $N-1$ dimensional surface, we conveniently choose the equation of
surface $f(\mathbf{x})=0$ such that $\left\vert \nabla f(\mathbf{x}%
)\right\vert =1$ so the normal $\mathbf{n\equiv}\nabla f(\mathbf{x})$ at a
local point $u^{%
\mu
}$, and $g_{\mu\nu}\equiv\partial\mathbf{x/\partial}u^{%
\mu
}\cdot\partial\mathbf{x/\partial}u^{\nu}=\mathbf{x}_{,\mu}\cdot\mathbf{x}%
_{,\nu}$ where $O_{,\mu}\equiv\partial O/\partial u^{\mu}$ and $O^{,\mu
}=g^{\mu\nu}O_{,v}$ etc. For a particle constrained on the surface, we have a
compatible constrained condition \cite{weinberg,jpn3,choice2,choice247},%
\begin{equation}
\mathbf{n\cdot p}=0. \label{0th}%
\end{equation}
The surface first category of FCR is \cite{weinberg,jpn3,choice2,choice247}:
\begin{equation}
\lbrack x_{i},x_{j}]=0,\text{ }[x_{i},p_{j}]=i\hbar(\delta_{ij}-n_{i}%
n_{j}),\text{ }[p_{i},p_{j}]=-i\hbar\left\{  ({n_{i}}{n_{k}}_{,j}-{n_{j}%
}{n_{k}}_{,i})p_{k}\right\}  _{Hermitian}, \label{1st}%
\end{equation}
where $O_{_{Hermitian}}$ stands for a suitable construction of the Hermitian
operator of an observable $O$. Because the classical Hamiltonian takes form
$H=p^{2}/2m+V$, we have \cite{jpn3} $[\mathbf{x},H]_{D}=\mathbf{p}/m$, and
$[\mathbf{p},H]_{D}=-\mathbf{n}{n_{k}}_{,j}p_{k}p_{j}$. The second category of
FCR is then given by,%
\begin{equation}
\frac{\mathbf{p}}{m}=\frac{i}{\hbar}[H,\mathbf{x}],\text{ }[H,\mathbf{p}%
]=\frac{i\hbar}{m}\left(  \mathbf{n}{n_{k}}_{,j}p_{k}p_{j}\right)
_{Hermitian}. \label{2nd}%
\end{equation}

Once curved surface $f(\mathbf{x})=0$ becomes flat, both the momentum and the
Hamiltonian must assume their usual forms respectively. Thus we ansatz that
the quantum mechanics $H$ takes the following form with a potential $V_{g}$,%
\begin{equation}
H=-\hbar^{2}/(2m)\nabla_{LB}^{2}+V+V_{g}. \label{ansatz}%
\end{equation}

\section{Geometric momentum}

First of all, we have the momentum ${\mathbf{p}}$ from Eqs. (\ref{2nd}) and
(\ref{ansatz}),%
\begin{equation}
{\mathbf{p}}=\frac{i}{\hbar}\frac{{{\hbar^{2}}}}{2}[\nabla_{LB}^{2}%
,{\mathbf{x}}]=-i\frac{\hbar}{2}\left(  {\left(  \nabla_{LB}^{2}{{\mathbf{x}}%
}\right)  +2{{\mathbf{x}}^{,\mu}}{\partial_{\mu}}}\right)  =-i\hbar
({\nabla_{S}}+\frac{{M{\mathbf{n}}}}{2}). \label{GM}%
\end{equation}
We call ${\mathbf{p}}$ (\ref{GM}) the \textit{geometric momentum} for its
dependence on the extrinsic curvature ${M}$
\cite{NYgeometry,ESgeometry,JNgeometry,geometry}.

Second, we demonstrate that the operator version of the constrained condition
(\ref{0th}) for the momentum ${\mathbf{p}}$,
\begin{equation}
\mathbf{p}\cdot\mathbf{n+n}\cdot\mathbf{p}=0. \label{vert}%
\end{equation}
It is evident for the action of the vector operator ${\nabla_{S}}$ on the unit
normal vector $\mathbf{n}$\ leads to a nonvanishing result as ${\nabla_{S}%
}\cdot\mathbf{n}=-M$, which exactly cancels $M$ in $\mathbf{(}{\nabla_{S}%
}+M\mathbf{\mathbf{n}}/2\mathbf{)}\cdot\mathbf{n+n}\cdot\mathbf{(}{\nabla_{S}%
}+M\mathbf{\mathbf{n}}/2\mathbf{)}=0$ so that we have orthogonal relation
(\ref{vert}).

Lastly, we need to show that the first category of the FCR are satisfied with
this momentum ${\mathbf{p}}$ (\ref{GM}). A verification of the first two FCR
in (\ref{1st}) is straightforward. The proof of the last in (\ref{1st}) is
also an easy task. The key step is a proper construction of the Hermitian
operator of observable $O$. Only the following naive rule $(O+O^{\dag})/2$
\cite{operator} is used,
\begin{equation}
\left\{  {n_{i}}{n_{k}}_{,j}p_{k}\right\}  _{Hermitian}=\frac{1}{2}\left(
{n_{i}}{n_{k}}_{,j}p_{k}+p_{k}{n_{i}}{n_{k}}_{,j}\right)  ={n_{i}}{n_{k}}%
_{,j}p_{k}+\frac{1}{2}(-i\hbar)\left(  {n_{i}}_{,k}{n_{j}}{_{,}{}_{k}}%
+{n_{i}\partial}_{{j}}M\right)  \text{.} \label{RHS}%
\end{equation}
It is applicable in the R.H.S. of the FCR $[p_{i},p_{j}]=-i\hbar\left\{
({n_{i}}{n_{k}}_{,j}-{n_{j}}{n_{k}}_{,i})p_{k}\right\}  _{Hermitian}$ as,%
\begin{equation}
\left\{  ({n_{i}}{n_{k}}_{,j}-{n_{j}}{n_{k}}_{,i})p_{k}\right\}
_{Hermitian}=\left\{  {n_{i}}{n_{k}}_{,j}p_{k}\right\}  _{Hermitian}%
-(i\longleftrightarrow j). \label{RHS2}%
\end{equation}

For a two-dimensional spherical surface, how to measure the geometric momentum
(\ref{GM}) is extensively investigated \cite{qhliu}.

\section{Geometric potential for quantum motion on spherical surface}

With the first category of the FCR being used only, we have at least ten
choices of $\alpha(N)$ in $V_{g}=\alpha(N)\hbar^{2}/(2mr^{2})$, based upon
various understanding of the problem. For instance, on the dependence of
$V_{g}$ on the dimensions $N$, we have, 1) $\alpha(N)=0$
\cite{Kleinert,choice1}, 2) $\alpha(N)=(N-1)^{2}/4$ \cite{choice247}, 3)
$\alpha(N)=(1+4s^{2})(N-1)^{2}/4$ with $s$ being a real parameter \cite{jpn2},
4) $\alpha(N)=N^{2}/4$ \cite{choice3}, 5) $\alpha(N)=(N-1)(N+1)/4$
\cite{choice247}, 6) $\alpha(N)=(N-1)N/4$ \cite{choice56}, 7) $\alpha
(N)=(N-3)(N+1)/4$ \cite{choice6}, 8) $\alpha(N)=(N-1)(N-2)\beta$
\cite{choice247} 9) $\alpha(N)$ arbitrary \cite{jpn2} and 10) $\alpha
(N)=(N-1)(N-3)/4$ \cite{jpn2}, etc. \cite{choice89} At first sight, these
disputant results seem to be rather irrelevant. Common experiments are only
capable of detecting energy differences, in which these constants drop out.
Cosmology, however, is sensitive to an additive constant
\cite{weinberg,Kleinert}.

Now let us see what $V_{g}$ is within our SCQS. The first category of the FCR
is \cite{Kleinert,choice247},
\begin{equation}
\lbrack x_{i},x_{j}]=0,[x_{i},p_{j}]=i\hbar(\delta_{ij}-n_{i}n_{j}%
),[p_{i},p_{j}]=-i\hbar(x_{i}p_{j}-x_{j}p_{i})/r^{2}, \label{sphere1}%
\end{equation}
No operator ordering problem occurs in the R.H.S. of $[p_{i},p_{j}]$ because
of the Jacobi identity. We see already these relations (\ref{sphere1}) are
automatically satisfied with Cartesian coordinates $\mathbf{x}$ and geometric
momentum $\mathbf{p}$ (\ref{GM}). Now we examine the remaining FCR in the
second category (\ref{2nd}) which is given by, with noting relations
$\mathbf{n}=\mathbf{x/}r$ and ${n_{i}}_{,j}=\left(  \delta{_{ij}}-n{_{i}%
}n{_{j}}\right)  /r$ so ${n_{i}}_{,j}p_{k}p_{j}=p^{2}/r=2mH/r$,%
\begin{equation}
\lbrack H,\mathbf{p}]=i\hbar\frac{\mathbf{x}H+H\mathbf{x}}{r^{2}}.
\label{sphere2}%
\end{equation}
On one hand, because the geometric potential $V_{g}$ results from the
noncommutability of different components of the geometric momentum (\ref{GM}),
it depends solely on the geometric invariants as the geometric momentum does.
On the other, all principal curvatures\ for the spherical surface are the same
$-1/r$ and the mean curvature is $M=-(N-1)/r$. So, the geometric potential
$V_{g}$ also takes the following form $\alpha(N)\hbar^{2}/(2mr^{2})$ and the
Hamiltonian takes form $H=-\hbar^{2}/(2m)\nabla_{LB}^{2}+\alpha(N)\hbar
^{2}/(2mr^{2})$. We will prove,
\begin{equation}
V_{g}=\frac{(N-1)(N-3)}{4}\frac{\hbar^{2}}{2mr^{2}}, \label{Vg}%
\end{equation}
which is exactly the geometric potential $V_{g}$ (\ref{GP}) for the surface
under consideration. The proof is as what follows.

We rewrite both $H$ into following form,%
\begin{equation}
H=-\frac{\hbar^{2}}{2m}\nabla_{LB}^{2}+V_{g}=\frac{\mathbf{p}^{2}}{2m}%
-\frac{M^{2}\hbar^{2}}{8m}+V_{g},
\end{equation}
and the quantity $(\mathbf{x}H+H\mathbf{x)}$ in the R.H.S. of the FCR
(\ref{sphere2}),%
\begin{equation}
\mathbf{x}H+H\mathbf{x}=2\mathbf{x}H-i\frac{\hbar}{m}\mathbf{p.} \label{RH}%
\end{equation}
The L.H.S. of the (\ref{sphere2}) $[H,\mathbf{p}]=[\mathbf{p}^{2}%
,\mathbf{p}]/(2m)$ is, with repeated use of the first category of FCR
(\ref{sphere1}),
\begin{equation}
\frac{1}{2m}[\mathbf{p}^{2},\mathbf{p}]=\frac{1}{{{r^{2}}}}\left(
{2i\hbar\mathbf{x}}\frac{{{p^{2}}}}{{2m}}-{2i\hbar\mathbf{x}}\frac{{{M^{2}%
}{\hbar^{2}}}}{{8m}}+{2i\hbar\mathbf{x}}{V_{g}}+{\frac{{{\hbar^{2}}}}%
{m}\mathbf{p}}\right)  {\mathbf{.}} \label{LH}%
\end{equation}
Multiplying the results (\ref{RH}) and (\ref{LH}) derived from both sides of
(\ref{sphere2}) by\ the unit normal vector $\mathbf{n}$ from the left, we
obtain (\ref{Vg}). \textit{Q.E.D.}

It is interesting to point out that the geometric momentum $\mathbf{p}$ and
the angular momentum $L_{ij}\equiv x_{i}p_{j}-x_{j}p_{i}$ to form a closed
$so(N,1)$ algebra. Let $P_{i}\equiv rp_{i}$, we have from (\ref{sphere1}),
\begin{equation}
\lbrack P_{i},P_{j}]=-i\hbar L_{ij}.
\end{equation}
It is easily to show that the components of the angular momentum satisfies the
standard $so(N)$ algebra from its definition of $L_{ij}$ and FCR
(\ref{sphere1}) \cite{weinberg},
\begin{equation}
\lbrack{L_{ij}},{L_{k\ell}}]=-i\hbar\left(  {-{\delta_{i\ell}}{L_{kj}%
+{\delta_{ik}}{L_{\ell j}}+\delta_{jk}}{L_{i\ell}}-{\delta_{j\ell}}{L_{ik}}%
}\right)  .
\end{equation}
The commutation relations between ${{L_{ij}}}$ {{and }}${{P_{\ell}}}$ is, with
also repeated use of the first category of FCR (\ref{sphere1}),
\begin{equation}
\left[  {{L_{ij}},{P_{\ell}}}\right]  =i\hbar\left(  {{\delta_{i\ell}}{P_{j}%
}-{\delta_{j\ell}}{P_{i}}}\right)  .
\end{equation}
These generators ${{L_{ij}}}$ and ${{P_{\ell}}}$ form a closed $so(N,1)$
algebra, which reflects a dynamical $SO(N,1)$ group symmetry beyond its
geometrical one $SO(N)$. It implies that there is a dynamical representation
that can be used to examine the motion on the spherical surface, as
illustrated in \cite{qhliu}.

\section{Conclusions and remarks}

The usual canonical quantization procedure contains only the first category of
the FCR, forming the invariable part of the procedure, therefore universally
valid. As widely accepted, this procedure is far from complete, and has free
parameters that are sometimes believed to be fixed by the experiments. For a
system that has classical analogue whose classical Hamiltonian takes form
$H=p^{2}/2m+V$, the freedom can be fixed by an additional principle that the
Cartesian coordinates must be used in passing over to the quantum mechanics
such that the second category of the FCR is automatically satisfied.

For a particle constrained to remain on a curved hypersurface, the different
components of momentum are not mutually commutable. Thus the procedure of
obtaining the quantum Hamiltonian by a simple substitution of an expression of
the momentum into the Hamiltonian $H=p^{2}/2m+V$ has been an issue full of
debates, and is therefore questionable. A further strengthening of the
quantization procedure is needed, and we propose to use the second category of
the FCR to determine the forms of both the quantum momentum and Hamiltonian.
The present study shows that there is a universal form of the momentum, the
geometric momentum, and there is a concise and lucid way to produce the
geometric potential for the spherical surface. Moreover, we demonstrate that
there is a dynamical $SO(N,1)$ group symmetry on the surface beyond the
geometrical one $SO(N)$.

There are interesting issues which will be explored in near future: the
relation between the geometric momentum and annihilation operators on the
$N-1$ dimensional sphere \cite{liu13}, and a possibly universal form of a
construction of the operator $\left(  \mathbf{n}{n_{k}}_{,j}p_{k}p_{j}\right)
_{Hermitian}$(\ref{2nd}) rather than a treatment on the case-by-case basis
\cite{arXiv1,arXiv2}, and the possible influence of the geometric potential on
the dark energy as a consequence of embedding our universe in higher
dimensional flat space-time, etc.

\begin{acknowledgments}
This work is financially supported by National Natural Science Foundation of
China under Grant No. 11175063. The author is grateful to Professor M.
Ritor\'{e}, Facultad de Ciencias, Universidad de Granada, Spain, for his kind
help of the differential geometry of hypersurface.
\end{acknowledgments}


\begin{thebibliography}{99}                                                                                               %


\bibitem {weinberg}S. Weinberg, \textit{Lectures on Quantum Mechanics},
(Cambridge University Press, Cambridge, 2013).

\bibitem {jk}H. Jensen and H. Koppe, Ann. Phys. \textbf{63}, 586(1971).

\bibitem {dacosta}R. C. T. da Costa, Phys. Rev. A \textbf{23}, 1982(1981).

\bibitem {jpn1}T. Homma, T. Inamoto, T. Miyazaki, Phys. Rev. D \textbf{42}, 2049(1990).

\bibitem {jpn2}M. Ikegami and Y. Nagaoka, Prog. Theoret. Phys. Suppl.
\textbf{106, }235(1991).

\bibitem {jpn3}M. Ikegami and Y. Nagaoka, S. Takagi and T. Tanzawa, Prog.
Theoret. Phys. \textbf{88, }229(1992).

\bibitem {liu07}Q. H. Liu, C. L. Tong and M. M. Lai, J. Phys. A: Math. and
Theor. \textbf{40}, 4161(2007).

\bibitem {EM}G. Ferrari and G. Cuoghi, Phys. Rev. Lett. \textbf{100}(2008)230403.

\bibitem {JD}B. Jensen and R. Dandoloff, Phys. Rev. A \textbf{80}, 052109 (2009).

\bibitem {OB}C. Ortix and J. van den Brink, Phys. Rev. B \textbf{81}, 165419
(2010); Phys. Rev. B \textbf{83}, 113406 (2011).

\bibitem {SOintl}M. V. Entin and L. I. Magarill, Phys. Rev. B \textbf{64},
085330(2001). A.V. Chaplik and L. I. Magarill, Phys. Rev. Lett. \textbf{96},
126402 (2006). M. P. L\'{o}pez-Sancho and M. C. Mu\~{n}oz, Phys. Rev. B
\textbf{83}, 075406 (2011).

\bibitem {liu11}Q. H. Liu, L. H. Tang, D. M. Xun, Phys. Rev. A \textbf{84, }042101(2011).

\bibitem {exp1}A. Szameit, et. al, Phys. Rev. Lett. \textbf{104}, 150403(2010).

\bibitem {exp2}J. Onoe, T. Ito, H. Shima, H. Yoshioka and S. Kimura, Europhys.
Lett. \textbf{98},\textbf{ }27001(2012).

\bibitem {dirac1}P. A. M. Dirac, \textit{The Principles of Quantum Mechanics},
4th ed. (Oxford Univ., Oxford, 1967).

\bibitem {dirac2}P. A. M. Dirac, \textit{Lectures on quantum mechanics}
(Yeshiva Univ., N. Y., 1964); Can. J. Math. \textbf{2}, 129(1950).

\bibitem {Klauder}J. R. Klauder, \textit{Quantization of Constrained Systems},
in \textit{Methods of Quantization}, ed. by H. Latal, W. Schweiger, Lecture
Notes in Physics Vol. 572 (Springer, Berlin, 2001) pp. 143-182.

\bibitem {note}Remarks on the terminology geometric potential: It was firstly
introduced in A. V. Chaplik and R. H. Blick, New J. Phys. 6, 33(2004). It has
different names in different papers, for instance, the embedding potential in
S. Matsutani, J. Phys. A: Math. Gen. 26, 5133(1993); the curvature potential
in M. Encinosa and L. Mott, Phys. Rev. A 68, 014102 (2003); the quantum
potential in P. Maraner, Ann. Phys. 246, 325 (1996); the effective
geometry-induced quantum potential in V. Atanasov, R. Dandoloff, A. Saxena,
Phys. Rev. B 79, 033404(2009); and the curvature-induced effective potential
in H. Shima, H. Yoshioka, and J. Onoe, Phys. Rev. B 79, 201401(R) (2009), etc.
I think that geometric potential and embedding potential are two appropriate names.

\bibitem {NYgeometry}I. Chavel, \textit{Riemannian Geometry: A Modern
Introduction}, 2nd ed. (Cambridge Univ., Cambridge, 2006) Exercise II.10, P.100.

\bibitem {ESgeometry}M. Ritor\'{e} and C. Sinestrari, \textit{Mean Curvature
Flow and Isoperimetric Inequalities}, ed. by V. Miquel, J. Porti
(Birkh\"{a}user Verlag, Berlin, 2010) Eq. (2.6) of P. 8.

\bibitem {JNgeometry}M. Kimura, in \textit{Topics in Mathematical Modeling},
Jind\v{r}ich N\v{e}cas Center For Mathematical Modeling, Lecture Notes, Vol.
4, ed. by M. Bene\v{s} and E. Feireisl (Matfyzpress, Prague, 2008), Lemma 2.1
of P. 53, Definitions 2.2-2.4 of P. 53, Corollary 2.12 of P. 55, and Theorem
2.17 of P. 56.

\bibitem {geometry}T. Frankel, \textit{The geometry of Physics: An
Introduction} (Cambridge Univ., Cambridge, 2004). p.224, p.305, p.310-311.

\bibitem {1925}P. A. M. Dirac, Proc. R. Soc. Lond. A \textbf{109}, 642(1925).

\bibitem {Schiff}L. I. Schiff, \textit{Quantum Mechanics}, (McGraw-Hill, New
York, 1949) p. 135.

\bibitem {Greiner}W. Greiner, \textit{Quantum Mechanics: An Introduction},
4th. ed. (Springer, Berlin, 2001) pp. 193--196.

\bibitem {essen}H. Ess\'{e}n, Am. J. Phys. \textbf{46, }983(1978).

\bibitem {arXiv1}D. M. Xun, Q. H. Liu, X. M. Zhu, e-print arXiv:1212.6373v2.

\bibitem {arXiv2}D. M. Xun, Q. H. Liu, e-print arXiv:1303.0909v1.

\bibitem {choice2}A. V. Golovnev, Rep. Math. Phys. \textbf{64,} 59--77(2009).

\bibitem {choice247}J. R. Klauder, S. V. Shabanov, Nucl. Phys. B \textbf{511,} 713--736(1998).

\bibitem {operator}J. R. Shewell, Am. J. Phys. \textbf{27,} 16(1959).

\bibitem {qhliu}Q. H. Liu, e-print arXiv:1209.2209v3.

\bibitem {Kleinert}H. Kleinert, S. V. Shabanov, Phys. Lett. A \textbf{232,} 327--332(1997).

\bibitem {choice1}B. Podolsky, Phys. Rev. \textbf{32,} 812--816(1928); P.
Dita, Phys. Rev. A \textbf{56,} 2574--2578(1997); C. Neves and C. Wotzasek, J.
Phys. A: Math. Gen. \textbf{33,} 6447--6456(2000).

\bibitem {choice3}A. G. Nuramatov and L. V. Prokhorov, Int. J. Geom. Meth.
Mod. Phys. \textbf{3, }1459--1467(2006)

\bibitem {choice56}J. A. Neto and W. Oliveira, Int. J. Mod. Phys. A
\textbf{14, }3699--3713(1999).

\bibitem {choice6}S.-T. Hong and K. D. Rothe, Ann. Phys. \textbf{311, }417--430(2004).

\bibitem {choice89}H. Grundling, C. A. Hurst, J. Math. Phys. \textbf{39,
}3091--3119(1998), and R. Froese, I. Herbst, Comm. Math. Phys. \textbf{220, }489--535(2001).

\bibitem {liu13}Q. H. Liu, Y. Shen, D. M. Xun, X. Wang, Int. J. Geom. Meth.
Mod. Phys. \textbf{10, }1320007(2006), and references cited therein.
\end{thebibliography}
\end{document}